\documentstyle[12pt]{article}
\begin{document}
\begin{titlepage}
\begin{center}

\null
\vskip-1truecm
\rightline{IC/94/280}
\vskip1truecm
{International Atomic Energy Agency\\
and\\
United Nations Educational Scientific and Cultural Organization\\
\medskip
INTERNATIONAL CENTRE FOR THEORETICAL PHYSICS\\}
\vskip3truecm
{\bf SUPERSYMMETRIC MAGNETIC MOMENTS SUM RULES\\
AND SPONTANEOUS SUPERSYMMETRY BREAKING\\}
\vskip2truecm
{S.S. Khalil \footnote{\normalsize Permanent address:
Department of Mathematics, Faculty of Science, Ain Shams University, Cairo,
Egypt.}\\
International Centre for Theoretical Physics, Trieste, Italy.\\}
\end{center}
\vskip1truecm

\baselineskip=24pt

\centerline{ABSTRACT}
\bigskip

In supersymmetry the anomalous magnetic moment of particles belonging to the
same supermultiplet is related by simple sum rules. We study the modification
of these sum rules in the case of the spontaneously broken N=1 global
supersymmetry.

\vskip-1truecm
\begin{center}
{MIRAMARE -- TRIESTE\\
\medskip
September 1994\\}
\end{center}
\end{titlepage}
\baselineskip=24pt

\section{Introduction}
It was shown [1] that N=1 supersymmetry gives rise to model independent sum
rules relating the magnetic transitions between states of different spin with a
given charged massive multiplet of arbitrary spin. Denoting the gyromagnetic
ratio $ g_j$ of a given spin j particle as defined by
$$\mu\!\!\!\! \mu_j = \frac{e}{2 M} g_j {\bf J} $$
These rules reduce to $g_{1/2} = 2$ for chiral multiplets and to
\begin{equation}
g_{1/2} = 2+ 2 h , \qquad g_{1}= 2 + h
\end{equation}
for vector multiplets, where h is a real number characterizing the magnetic
transition between the spin-0 and spin-1 states.

The relevant question for phenomenological implication is how these sum rules
modify in the physical situation where the supersymmetry is broken. A first
attempt in this direction was accomplished in the work of Ref.[2] where a
class of
soft supersymmetry breaking terms were introduced.

In this paper we would like to study the modification of these rules with a
theory
in which supersymmetry is spontaneously broken. The hope is that the
spontaneous nature of supersymmetry breaking can guarantee the survival of
some interesting relations among the various transition magnetic moments. This
can be seen as an intermediate step for a full comprehension of the problem
in a
theory where a local supersymmetry is spontaneously broken to a global N = 1
SUSY theory plus a set of SUSY breaking terms.

The most general CP and $U(1)_{e.m.}$ invariant $WW\gamma $ vertex when all
particles are on mass shell is [3]
\begin{equation}
M_{\mu \alpha \beta}=ie\{A[2p_{\mu}g_{\alpha \beta}+4(q_{\alpha}g_{\beta
\mu}-
q_{\beta}g_{\mu \alpha})]+2\Delta K_{WW}(q_{\alpha}g_{\beta \mu}-q_{\beta}
g_{\mu \alpha})+4\frac{\Delta Q}{m^2_W} p_{\mu}q_{\alpha}q_{\beta}\}
\end{equation}
where $p-q$, $p+q$, $2q$ are the momenta of the incoming and outgoing $W^+$
and of
the incoming photon. At the tree level of the Minimal Supersymmetric Standard
Model A=1,  the anomalous magnetic moments and the electric quadruple
moment $\Delta K_{WW}$ and $\Delta Q$ are equal to zero. This agrees with the
fact that in a renormalizable theory of spin $ 1\over 2 $ and spin 1
particles, the
tree level value of the gyromagnetic ratio is equal to 2 and this is
implied by the
tree level unitarity [4,5].

Radiative contributions modify these tree level results. In the
supersymmetric case it
was shown that the existing sum rule holds at any order in perturbation theory
implying that:
\begin{equation}
\Delta K_{WW}=a_{\omega_1}=a_{\omega_2}=\Delta K_{WH}. \end{equation}
where $ a_{\omega_1}$ and $a_{\omega_2}$ are the anomalous magnetic
moments of the charginos $w_1$ and $w_2$, $ a_{\omega_i}= \frac{ g_{w_i}-
2}{2}$. They are given by the coefficient of $$\frac{1}{2 m_{w_i}} \bar{w_i}
\sigma^{\mu \nu} q_{\nu} w_i \varepsilon_{\mu}$$ where $ e_{w_1} = - e $ , $
e_{w_2} = + e $, q and $\varepsilon$ are the momentum and the polarization
vector of the incoming photon.

$\Delta K_{WH}$ is the magnetic transition between the spin 1 and spin 0
state in
the vector multiplet. It is characterized by the coefficient of $$
\frac{e}{m_W}
\varepsilon^{\mu \nu \rho \sigma} p_{\rho} q_{\sigma} \varepsilon_{\mu}
\varepsilon_{\nu}' $$ where p, $\varepsilon_{\nu}'$ , q and $\varepsilon_{\mu}$
are the momentum and polarization vector of the incoming $ W^+$ and $\gamma$
respectively.

A first explicit verification of these rules can be found in Ref.[6] where the
equality of $\Delta K_{WW}$ and $ a_{\omega_i}$ was shown in the case of
massless ordinary fermions. As for massive fermions two cases have been
studied:

1- $ m_{t} >> m_{W}$, with fixed
ratio $ ( \frac{ m_b}{m_t} )^2= r $. In this case the supersymmetric sum
rules are
satisfied [7] and
\begin{equation}
\Delta K_{WW}^{ql}=a_{\omega_{1}}^{ql}=a_{\omega_{2}}^{ql}=\Delta K_{WH}^{ql}=
\frac{-g^2}{32\pi^2} F(r) \label{dkql}
\end{equation}
with
\begin{equation}
F(r)=
\frac{1}{(1-r)^3} [r^3+11r^2-13r+1-4r(1+2r)\ln{r} ]. \end{equation}
2- $ m_b=0$, with fixed ratio $( \frac{m_W}{m_t})^2 = \alpha$. In this case we
have [8]
\begin{equation}
\Delta K_{WW}^{q\tilde{q} l\tilde{l}} = a_{\omega_1}^{q\tilde{q} l\tilde{l}} =
a_{\omega_2}^{q\tilde{q} l\tilde{l}} = \Delta K_{WH}^{q\tilde{q} l\tilde{l}} =
\frac{-g^2}{32\pi^2} G(\alpha )
\end{equation}
with
\begin{equation}
G(\alpha )=\frac{2}{\alpha^2} [3\alpha +(3-2\alpha )\ln (1-\alpha )].
\end{equation}

In the Minimal Supersymmetric Standard Model (MSSM) with supersymmetry
broken explicitly but softly by a universal mass $\tilde{m}$ for all scalar
particles, the total contribution of the four quantities has been
considered [2].
However, in this case, the sum rule (3) results to be badly broken without any
interesting functional relation among the four quantities in  Eq.(3). In
this note
we intend to pursue the same strategy in the case of spontaneous breaking
of the
global N=1 supersymmetry.

For the sake of argument we will consider a SUSY spontaneous breaking realized
a`la Fayet-Iliopoulos in the realization of Ref.[8]. In that model the
following mass
splitting is obtained:
\begin{eqnarray}
\Delta m_{gauge}^2 = \mu^2 , \\
\Delta m_{matter}^2 = {1\over 4} \mu^2
\end{eqnarray}
where $\Delta m_{gauge}$ and $\Delta m_{matter}$ denote the mass difference
between the fermionic and bosonic components in the vector and scalar multiplet
respectively, and $\mu$ is the proportional to the v.e.v of the
Fayet-Iliopoulos
term.

The outline of the paper is as follows. In Section 2, we present the
calculation of
the anomalous magnetic moment of the muon in the simplified case of
spontaneously broken super QED. In Sections 3 and 4, we calculate the anomalous
magnetic
moment of the spin 1 -gauge boson and spin $ 1\over 2 $ partner respectively.
In Section 5, we discuss the modification of the supersymmetric sum rules
in this
model, and present our conclusion.

\section{The anomalous magnetic moment of the muon}

The anomalous magnetic moment of the muon, $ a_{\mu}$, is one of the most
precisely known quantities, and, hence, it can be used to look for the physics
beyond the standard model. In N=1 global super QED this quantity was shown to
be vanishing at any order in perturbation theory [9].

Here we compute the anomalous magnetic moment of the muon in super QED
with spontaneous breaking a`la Fayet, for definiteness we consider the mass
splitting of Eq.(8). The graphs which will contribute to the anomalous magnetic
moment of $ a_{\mu}$ are given in Fig.1.

The first diagram, where the muon and photon are running in the loop, gives a
contribution to $ a_{\mu}$ equal to $ \frac{\alpha}{2\pi}$.

 From the second diagram, where smuon and photino are running in the loop, one
obtains:
B\begin{equation}
\frac{\alpha}{\pi} \int_0^1 dx\, \frac{x(1-x)(a^2x+a)}{-a^2x^2 +3/4 x -1}
\end{equation}
where $ a^2 = \frac{m_{\mu}^2}{\mu^2}$ and two cases arise:

When $ a \rightarrow 0 $, i.e $ \mu^2 \gg m_{\mu}^2$, i.e supersymmetry is
broken at a high scale, we find that Eq.(9) gives zero, and the anomalous
magnetic moment is given by $ \frac{\alpha}{2\pi}$, we recover the QED result.

When $ a \rightarrow \infty $ , i.e $\mu^2 \rightarrow 0$ i.e supersymmetry is
unbroken. Eq.(9) gives a value equal to $ - \frac{ \alpha}{2\pi}$ which cancels
the
contribution of the first diagram. Then the anomalous magnetic moment is $
a_{\mu}= 0 $ , in agreement with the theorem of Ref.[9].

\section{The anomalous magnetic moment of the W-boson}

We calculate the one loop contributions to the anomalous magnetic moment of
the $ W^+$-boson, $ \Delta K_{WW}$, with quark and lepton supermultiplet
running in the loop. We have a contribution from two diagrams shown in
Fig.2. The
results obtained from these graphs are in agreement with [2], provided one
considers for sfermion masses the SUSY breaking contribution of Eq.(8). For
instance in the top generation these results are
\begin{eqnarray}
\Delta K_{WW} (bbt) =\ \frac{g^2 N_c q_b}{32\pi^2} \int_0^1 dx\,
\frac{x^4+x^3(a-
b-1)+x^2(2b-a)}{bx+a(1-x)-x(1-x)} \\ \Delta K_{WW} (ttb) = - \frac{g^2 N_c
q_t}{32\pi^2} \int_0^1 dx\,
\frac{x^4+x^3(b-a-1)+x^2(2a-b)}{ax+b(1-x)-x(1-x)} \\
\Delta K_{WW} (\tilde b \tilde b \tilde t) = - \frac{g^2 N_c q_b}{16\pi^2}
\int_0^1
dx\, \frac{(x^3-x^2)(\tilde b - \tilde a -1 +2x)}{\tilde b x + \tilde a
(1-x) -x (1-x) }
\\
\Delta K_{WW} (\tilde t \tilde t \tilde b) = \ \frac{g^2 N_c q_b}{16\pi^2}
\int_0^1 dx\, \frac{(x^3-x^2)(\tilde a - \tilde b -1 +2x)}{\tilde a x +
A\tilde b (1-x) -
x (1-x) }
\end{eqnarray}
where $ a = \frac{m_t^2}{m_W^2}$, $b=\frac{m_b^2}{m_W^2}$, $\tilde a =
\frac{m_{\tilde t} ^2}{m_W^2}$, $ \tilde b = \frac{m_{\tilde b}^2 }{m_W^2}$
and $
N_c$ is the number of colours.

The contribution of the leptons of that generation can be obtained from
Eq.(10),
and of the slepton from Eq.(12). In the case of massless fermions the above
equations yield \begin{equation}
\Delta K_{WW}^{q \tilde q l \tilde l} = - \frac{4g^2}{16 \pi^2} \int_0^1
dx\, \frac{ f
x (2x -1)}{f - x(1-x)}
\end{equation}
where $f= \frac{1/4 \mu^2}{m_W^2 }$. We will examine the value of this integral
for two cases:\\

1- $\mu = M_W$, i.e f= 1/4 and we found $ \Delta K_{WW}^{q \tilde q l
\tilde l} =
- \frac{4g^2}{16 \pi^2} [ 1/2 - \frac{1}{4} \pi ]$.\\

2- $\mu >> M_W$ we found $ \Delta K_{WW}^{q \tilde q l \tilde l} = -
\frac{4g^2}{16 \pi^2} [ 1/6 ] $.

\section{The anomalous magnetic moment of the gauginos}

We calculate the one loop contribution of the spin 1/2 supersymmetric partners
of the W-bosons ``$\omega_1$, $\omega_2$". We have a contribution from four
diagrams shown in Fig.3. The results of these graphs are in agreement with [2]
when taking $ m_{w}^2 = m_{W}^2 + \mu^2$, and they are
\begin{eqnarray}
a_{\omega_1} (b\tilde t \tilde t) &={\displaystyle\frac{g^2 N_c q_t}{16\pi^2}
\int_0^1
dx\, \frac{x(x-1) [B(x-2)+x]}{\tilde A x+ B(1-x)-x(1-x)}} \\
a_{\omega_1} (b b \tilde t) &={\displaystyle\frac{g^2 N_c q_b}{16\pi^2}
\int_0^1
dx\, \frac{x^2 [B(x+1)+x-1]}{ B x+ \tilde A(1-x)-x(1-x)} }\\
a_{\omega_1} (t \tilde b \tilde b) &={\displaystyle\frac{g^2 N_c q_b}{16\pi^2}
\int_0^1
dx\, \frac{x^2\tilde B (1-x)}{\tilde B x+ A (1-x)-x(1-x)} }\\
a_{\omega_1} (t t \tilde b) &={\displaystyle\frac{g^2 N_c q_t}{16\pi^2}
\int_0^1
dx\, \frac{x^2 \tilde B (1-x)}{ A x+ \tilde B (1-x)-x(1-x)} }
\end{eqnarray}
where
\begin{eqnarray}
A = \frac { m_t^2}{m_{\omega_1}^2} = \frac{m_t^2}{m_W^2+\mu^2}\\ \tilde A =
\frac{m_{\tilde t} ^2}{m_{\omega_1}^2} = \frac{m_t^2 +1/4 \mu^2}
{m_W^2+\mu^2}\\
B = \frac { m_b^2}{m_{\omega_1}^2} = \frac{m_b^2}{m_W^2+\mu^2}\\ \tilde B =
\frac{m_{\tilde b} ^2}{m_{\omega_1}^2} = \frac{m_b^2 +1/4 \mu^2}
{m_W^2+\mu^2}
\end{eqnarray}
if we consider the case of the massless fermions, we find \begin{equation}
a_{\omega_1}^{q\tilde q l \tilde l}=a_{\omega_2}^{q\tilde q l \tilde l}=
\frac{ 2
g^2}{16 \pi^2} \int_0^1 dx\, \frac{(1+r)x(2x-1)}{r-x} \end{equation}
where $$ r = \frac{1/4 \mu^2}{m_W^2 + \mu^2} = \frac{f}{1+4f}$$
If we
examine the value of this integral for the two cases in the previous
section we find
$$ a_{\omega_1}^{q\tilde q l \tilde l}
= a_{\omega_2}^{q\tilde q l \tilde l}=
\frac{ 2 g^2}{16 \pi^2} [\frac{-27}{256} \pi + \frac{ 9(-8+ 3 log(7))}{256}
],$$\\ $$
a_{\omega_1}^{q\tilde q l \tilde l}=a_{\omega_2}^{q\tilde q l \tilde l}=
\frac{ 2 g^2}{16 \pi^2} [ \frac{-5}{32} \pi + \frac{ 5(-4 + log(3) )}{125}]$$
respectively.

\section{Concluding remarks}

In the case of massless fermions the quark and lepton supermultiplet
contribution to $ \Delta K_{WH}$ vanishes because of the zero Yukawa coupling.
Hence if we compare Eqs.(15), (24) and their results in the cases which we have
studied with the result for $\Delta K_{WH}$ in the massless case, it
appears that
the sum rule (3) is badly broken without any surviving clear pattern. For
example in the realistic case where $\mu >> M_W$ we found that $ \Delta
KB_{WW}^{q \tilde q l \tilde l} = - 0.041667 \frac{g^2}{\pi^2}$ and $
a_{\omega_1}^{q\tilde q l \tilde l}=a_{\omega_2}^{q\tilde q l \tilde l}= -
0.07632
\frac{g^2}{\pi^2}$. Obviously, considering Eqs.(11)-(14) and Eqs.(16)-(19),
it is
clear that the problem becomes even more complicated in the case of massive
fermion.

In conclusion, our analysis indicates that even the spontaneous breaking of
global
supersymmetry completely destroys the sum rule in Eq.(3) without leaving any
simple pattern.

The investigation of this problem in the fully realistic case of spontaneously
broken N=1 supergravity is left for future investigation.
\vskip2truecm

\centerline{{\bf Acknowledgments}}
\bigskip

The author would like to thank Professor A. Masiero for his guide  and A.
Culatti
for useful discussions.  He would also like to thank Professor Abdus Salam, the
International Atomic Energy Agency and UNESCO for hospitality at the
International Centre for Theoretical Physics, Trieste.

\newpage

\begin{Large}
\begin{bf}
\begin{tabbing}
References
\end{tabbing}
\end{bf}
\end{Large}
\begin{enumerate}

\item { S. Ferrara and M. Porrati}, Phys. Lett. {\bf B288} (1992) 85.

\item{ A. Culatti}, DFPD/94/TH/25 (1994).

\item { W. A. Bardeen, R. Gastmans and B. Lautrup}, {\bf B46} (1972) 319.

\item { S. Weinberg}, in Lectures on Elementary Particles and Quantum Field
Theory, Proc. of the Summer Institute, Brandeis University, 1970, ed. S. Deser
(MIT Press, Cambridge, MA, 1970), Vol. I.

\item { S. Ferrara, M. Porrati and V. Telegdi}, Phys. Rev. {\bf D46} (1992)
3529.

\item {Bilchak, R. Gastmans and A. van Proeyen}, Nucl. Phys. {\bf B273}
(1986) 46.

\item{ S. Ferrara and A. Masiero}, CERN TH-6846/93, Proc. of "SUSY 93"
Northeastern University, Boston, USA (1993) and Proc of 26th Workshop,
Eloisatron Project, ``From Superstring to Supergravity", Erice, 1992.

\item { R.Barbieri, S.Ferrara}, D.V.Nanopulos, Z.Phys. C (1982) 267.\\
R.Barbieri, S.Ferrara, D.V.Nanouplos, Phys. Lett {\bf B116} (1982) 16.

\item{ S. Ferrara and E. Remiddi}, Phys. Lett. {\bf B53} (1974) 347.

\end{enumerate}

\newpage

\centerline{FIGURE CAPTIONS}
\bigskip

\begin{description}
\item{Fig.1}
Two diagrams contributing to the anomalous magnetic moment of the muon.

\item{Fig.2}
Two diagrams contributing to the anomalous magnetic moment of the W--boson.

\item{Fig.3}
Four diagrams contributing to the anomalous magnetic moment of the gaugino.
\end{description}

\end{document}